\documentclass[12pt]{article}

\usepackage[utf8]{inputenc}
\usepackage[T1]{fontenc}
\usepackage{authblk} %Pour affiliatiaon
\usepackage{amsmath,amsfonts,amssymb,amsthm}
\numberwithin{equation}{section}
\usepackage{dsfont} %Pour math ds
\usepackage{color}
\usepackage[dvipsnames]{xcolor} % for more colours and to invent some
\colorlet{mygreen}{green!60!black}
\colorlet{mygrey}{white!60!black}
\usepackage{graphicx} % pour l'inclusion de figures en eps,pdf,jpg
\usepackage{float}  %Permet de placer la figure ou tu veux
\usepackage{booktabs} %Permet \toprule et \midrule dans l'env. tabular
\usepackage{setspace} %Pour changer l'interligne
\usepackage{multicol} %Pour des équations a plusieurs col.
\usepackage{url} 
\usepackage[dvipsnames]{xcolor}
\usepackage[
  colorlinks = true,
  linkcolor  = MidnightBlue,  % equations, sections, figures
  citecolor  = MidnightBlue,  % bibliography citations
  urlcolor   = MidnightBlue   % URLs, DOIs
]{hyperref}

\usepackage[numbers,sort&compress]{natbib}

\usepackage{geometry} %Mise en page
\usepackage{palatino} %Fonts

\title{\huge Another Triumph of Locality \\[0.1em] 
\LARGE Colliding Histories Skew Handshakes\thanks{Published in \emph{Bold Conjectures, Volume II: Essays Across Physics}, edited by Logan Chipkin, Conjecture Press, 2026.}}

\author{Charles Alexandre B\'edard}
\affil{\small {École de technologie supérieure}\\
\footnotesize \emph{charles.alexandre.bedard@etsmtl.ca}}
\date{July 2026}

\begin{document}
\maketitle
\thispagestyle{empty} %Removes page number

%modifs for v5: modif

\begin{abstract}
\noindent From gravity to electromagnetism, apparent action at a distance has always been resolved by deeper, local explanations. Yet today, Bell's theorem is widely interpreted as the death knell for local reality. In this chapter, I present the theorem in accessible terms, examine the three main strategies that attempt to preserve hidden variables, and argue that they share a common defect: the attempt to explain the quantum from the classical rather than the other way around. 
%modif V4 a V4.1
When quantum mechanics is applied universally, classicality itself is given a quantum account; and when the Bell scenario is formulated in the Heisenberg picture, a strictly local explanation emerges. This chapter serves as a non-technical front-end to \emph{Explaining Bell Locally} (Proc. R. Soc. A).
\end{abstract}

\section{Bold Conjectures Against Action at a Distance}

Physical events can have their histories traced. We track the chain of causes link by link, across space and through time, until we reach the effect. If a link is missing—if a cause leaps across a void to produce an effect without crossing the space between them—then we do not have an explanation, we have a miracle. Physics has a history of eliminating miracles and replacing them with local mechanisms. Yet today, at the heart of quantum mechanics, we are told the miracle has returned. 

\emph{This resignation is premature.} 
In both gravity and electromagnetism, whenever we have faced the temptation of action at a distance, the solution has never been to abandon locality, but to save it.
Quantum theory is no exception.

Newtonian mechanics was first and foremost a bold conjecture of \emph{universality}. For centuries, the heavens and the Earth were assumed to operate under entirely different physical rules. Newton's radical claim was that nature is governed everywhere—from terrestrial projectiles to celestial orbits—by a single dynamical framework. The same equations that trace the fall of an object on Earth also govern the elliptical paths of the planets and the rhythm of the tides. This unification demonstrated that the cosmos is not fractured into distinct realms; it is bound by universal laws.

The very success of Newton's unification gave rise to a better problem: his law of gravity stood in tension with the principle of local mediation. While the mathematical formula worked with astonishing precision, it implied an instantaneous influence. Because the gravitational pull depended on the instantaneous distance between masses, the equation permitted no time delay: shifting matter \emph{here} would immediately change accelerations \emph{there}, leaving no room for a mediating mechanism.

Newton himself was haunted by this action at a distance, recognizing it not as a profound truth, but as a severe conceptual defect. He famously wrote: ``...that one body may act upon another at a distance through a vacuum without the mediation of anything else [\dots]
%by and through which their action and force may be conveyed from one to another 
is to me so great an absurdity that I believe no man who has in philosophical matters a competent faculty of thinking can ever fall into it''~\cite{newton_bentley_1693}.

Nearly a century after Newton, the same tension resurfaced when Coulomb formulated the law of electrostatic force. Here again, the equation worked spectacularly well, but because it shared the same mathematical structure as Newton's gravity, it resurrected the same conceptual defect. Since the force between charges depended on their distance at \emph{that instant}, their influence appeared to cross empty space without local propagation.

The problems of apparent instantaneity in gravity and in electricity looked like the same defect in two costumes. But no. They were not solved by the same idea, nor even in the same century. As it turns out, the younger problem was solved first.

Distrusting action at a distance, Faraday envisioned invisible ``lines of force'' filling the void between charges. Like wheat and oat fields that fill a landscape, electricity and magnetism were conjectured to have fields of their own that permeate space. What a charge does is to establish a local condition in its surroundings, and what another charge experiences is determined by the field at its own location. In place of distant tugging, Faraday proposed mediation through fields.

Maxwell supplied the decisive mathematical step by turning Faraday's picture into a rigorous set of local dynamical laws. Electric and magnetic fields became two aspects of a single electromagnetic field: a physical system spread through space, evolving point by point, carrying energy and momentum, and transmitting influence only at finite speed. What had once looked like instantaneous action at a distance was re-expressed as ordinary local dynamics in the intervening medium. Coulomb's law was therefore only an approximation: its apparent instantaneity was a fast-propagation limit, hiding the field's local mediation behind the sheer largeness of the speed of light.

Einstein's solution to the problem of gravity was deeper still. Gravity could not be understood as a force acting within a fixed stage of space and time, because while a force may deflect trajectories, it cannot make clock rates vary from place to place. Gravity does. It alters the very standards of space and time by which trajectories are measured, and once clocks and rulers become part of the phenomenon, the geometry they reveal for spacetime is not flat. It is curved, affected by mass and energy. The medium is therefore not a field living on spacetime; it is spacetime. What we call locality becomes the rule that influences propagate only within the limits set by the geometry of spacetime. Light, gravitational attraction and any other physical signal are bound to respect that structure.

These episodes are worth recalling for one simple reason: they show what it takes to keep locality. When a successful law looks like action at a distance, physics does not canonize the miracle. It demands a deeper, local explanation.

Bell's theorem is widely regarded as a case for which this strategy finally fails. 
Backed by decades of rigorous experiments, one often hears that quantum theory has taught us a bitter new fact: nonlocality is not a temporary embarrassment, but a fundamental feature of the world.
%``Quantum nonlocality'' has been baptized as a phenomenon.

To see why locality survives Bell's theorem, we must first strip away the inconsistent compromises of textbook orthodoxy. The illusion of nonlocality does not emerge from the laws of physics, but from a failure to apply those laws universally.

\section{Universality and the Return of Separability}

Quantum theory is the most accurate framework physics has ever produced. It underwrites the stability of atoms and chemistry, the spectra of stars, the behaviour of semiconductors and lasers, and the operating principles of today's clocks and computers. Across scales and platforms, its quantitative predictions are confirmed to absurd precision—not only in physics laboratories, but as the invisible infrastructure of modern technology. And yet this unrivalled success arrives with an awkward legacy: textbook quantum theory teaches an inconsistency that has become so entrenched that it is often mistaken for a feature.

On the one hand, quantum systems are said to obey a process called \emph{unitary evolution}, governed by the Schrödinger equation. This law is continuous, deterministic, reversible and local. Interactions occur through direct coupling of systems, and nothing in the unitary evolution resembles an instantaneous influence across space.

On the other hand, measurements are said to obey a collapse rule. Collapse is discontinuous, stochastic, irreversible and nonlocal: a measurement \emph{here} updates the state assigned to a distant system \emph{there}. The theory thus oscillates between two incompatible kinds of dynamics, while leaving the boundary between them—what counts as a ``measurement''—undefined.

In 1957, Hugh Everett III solved this inconsistency by denying the boundary~\cite{everett_universal_wavefunction_1973}.
Measurements, he insisted, are not a special category of event; they are physical interactions. 
Similarly, observers are not external classical entities; they are physical systems, governed by the exact same quantum theory as everything else.
After all, observers too are made of atoms.

The bold step here is often misdescribed. It is common to say that Everett's conjecture was ``many worlds''---that he replaced the collapse rule with a rule that multiplies the universes. But the central act was not to \emph{postulate} additional worlds. It was to take the Schrödinger equation to be universal---to treat unitary evolution as applying without exception, including to apparatus and observers.

Everett's proposal, therefore, belongs to the same family as Newton's universality: the insistence that the dynamical law does not come with an escape hatch. The multiplicity of worlds is not an added ingredient. It is what unitary evolution entails once one stops quarantining observers outside the dynamics.

%Old: Under unitary evolution, when an observer measures a quantum system in a superposition, she is swept up into it—smoothly branching, for instance, into a version who saw outcome~0 and a version who saw outcome~1. The ``observer-who-saw-0'' evolves alongside the system in state~0, establishing from within the overarching quantum state an autonomous, classical-like \emph{history}---experienced from within as a distinct \emph{world}. Meanwhile, the ``observer-who-saw-1'' follows another history.

Under unitary evolution, when an observer measures a quantum system in a superposition, she is swept up into it---smoothly branching, for instance, into a version who saw outcome~0 and a version who saw outcome~1. But she does not branch alone. The measurement apparatus and countless nearby systems are similarly swept up, becoming physical records of the different outcomes. 
The ``observer-who-saw-0'' evolves alongside this collection of witnesses to state~0, establishing from within the overarching quantum state a classical-like \emph{history}---experienced from the inside as a distinct \emph{world}. Meanwhile, the observer-who-saw-1 follows an entirely separate history.

%Very OLD% The "observer-who-saw-0" evolves alongside the system in state~0, establishing from within the overarching quantum state an autonomous, classical-like branch---a \emph{world}. And because she will never interact with her counterpart again, the "observer-who-saw-1" evolves in another world.

\sloppy
This starkly clashes with our everyday experience since we never feel branching nor are we ever aware of our counterparts. But we are used to such clashes. The Copernican revolution revealed that the Earth is a spinning ball, while Galilean and Newtonian mechanics explained exactly why we do not feel it moving. Similarly, quantum theory explains exactly what we observe: according to the laws of motion, the different versions of the observer cannot perceive or interact with one another. The histories are autonomous. Each simply looks around, sees a single definite measurement outcome, and feels perfectly normal.

Applying the Schrödinger equation without exception removes the most blatant source of nonlocality: collapse. But it does not yet fully settle the issue. For Einstein, locality was not merely the prohibition of faster-than-light effects; it was the stronger requirement that the real factual situation of a system \emph{here} be independent of what is done to a distant system \emph{there}~\cite{einstein_autobio_1949}. 

Because of a phenomenon known as \emph{entanglement}, this requirement cannot be fulfilled within the standard mathematical language of quantum mechanics.
In that standard language, known as the Schrödinger picture, entanglement dictates that a composite system cannot be described merely as a state of one part together with a state of the other. The theory assigns a single joint state to the pair. Because this single global state irreducibly describes two possibly remote systems, an operation performed \emph{there} instantly changes the very same state used to describe what is \emph{here}. We are left with an account in which the whole is primary, and the parts cannot carry their own local, complete state.

Yet historically, quantum theory was discovered twice in the span of a year, in two very different mathematical forms: Heisenberg's matrix mechanics and Schrödinger's wave mechanics. Because both frameworks yielded the exact same empirical predictions, they were quickly declared equivalent, and Schrödinger's approach became the orthodox language of physics. But demanding only identical predictions is a low bar for equivalence. When we ask what these theories say about the underlying reality—and in particular, what they say about locality—they tell profoundly different stories~\cite{bedard_realism_2026}.

As Deutsch and Hayden demonstrated~\cite{deutsch_hayden_2000}, Einstein's demand had been sitting in Heisenberg's formulation all along. 
%Returning to the Heisenberg picture provides a mode of description complete enough to recover all empirical predictions, yet strictly local.
%
In this picture, each subsystem carries its own strictly local ``descriptor''—a matrix that remains unaffected by actions performed on remote systems.
Moreover, the description of a composite system is nothing more than the collection of the descriptors of the parts. This grants a separable, factorizable description of systems without denying entanglement.

Everett's original analysis was formulated in the Schrödinger picture, where the universal state decomposes into autonomous versions of a system, such as the observer-who-saw-0 and the observer-who-saw-1.
The Heisenberg-picture analogue is that the system's local descriptor can itself be refined into a collection of autonomous components~\cite{kuypers_deutsch_2021}, each encoding a definite outcome.

We are now equipped to face the challenge of Bell's theorem. The prevailing wisdom has long been that a local explanation is not just difficult, but impossible. As the physicist Nicolas Gisin famously put it: ``Quantum nonlocal correlations emerge, somehow, from outside space–time in the sense that no story in space–time can tell how they happen''~\cite{gisin_science_2009}.

\emph{This chapter is that story}. It is a non-technical companion to ``Explaining Bell Locally''~\cite{bedard_rspa_2025}, where the full formal development and additional details are worked out.

\section{Bell's Theorem as a Game}

Let us begin with a simple thought experiment, a cooperative game called \emph{Coordinated Hidden Strategy Heroes} (CHSH). The game's initials secretly nod to its originators: Clauser, Horne, Shimony and Holt~\cite{clauser_horne_shimony_holt_1969}, who turned Bell's theorem~\cite{bell_1964_epr} into a testable experimental protocol.
At first glance, the game looks too simple to carry any deep message about the structure of reality. Yet it does. 

\emph{Coordinated Hidden Strategy Heroes} is played cooperatively by Alice and Bob. Before the game begins, they meet to devise a plan to maximize their score. They can brainstorm, share notes and instructions, or exchange a prearranged sequence of random numbers if they deem it helpful. They write a Strategy Card to carry with them. This card is the physical embodiment of their coordination, containing every bit of knowledge and information they possess to beat the game. 

In this preparatory phase, they also review the terms of victory. For each round of the game, Alice and Bob each receive a binary question, 0 or 1, and each must return a binary answer,~0 or~1. The combination of their questions determines whether their answers must agree or disagree. If at least one question is 0—that is, if the question pair is~(0,\,0),~(0,\,1), or~(1,\,0)—Alice and Bob are required to return the same answer. If, however, the question pair is~(1,\,1), they must return different answers. That is the entire game. Simple, isn't it?

The preparatory phase ends as the players are separated across the solar system. Alice travels to Mercury, while Bob travels to a base orbiting Jupiter, placing them more than thirty light-minutes apart. Once in position, they each receive a list of 40 questions, determining 40 rounds of \emph{Coordinated Hidden Strategy Heroes}. After recording their answers, they reconvene on Earth to compare results and tally their total score out of 40.

At their respective stations, the players have only five minutes to answer the 40 questions. 
Across a divide of more than thirty light-minutes, this tight schedule turns the speed of light into a fundamental limit to communication:
no message sent during the game can reach the other station before time runs out. 
Their pre-arranged Strategy Card is all they have left to rely on, as they are now isolated by space itself.

Now put yourself in Alice’s place, in the preparatory phase, discussing strategy with Bob. Pause for a moment and consider how you would coordinate for the game.

A common approach is the \emph{All-Zero Strategy}: both you and Bob simply agree to answer~0 across all questions, regardless of what is asked. Since 0 always matches 0, this guarantees a win in three of the four scenarios:~(0,\,0),~(0,\,1) and~(1,\,0). You only fail in the~(1,\,1) case, which requires different answers. Assuming the questions are drawn at random, this simple plan secures a win rate of 75\%.

Can we do better? 

Let us try to write a perfect Strategy Card. Since the difficulty lies in the~(1,\,1) scenario—where the players must disagree—let us prioritize that case. To secure a win here, Alice and Bob must coordinate to give different answers. For instance, they might specify: \emph{If Alice is asked~1, she writes~0; if Bob is asked~1, he writes~1}. This guarantees success for~(1,\,1). However, as we shall see, this choice forces their hand for the other scenarios, turning their strategy into a logical armlock.

If the question pair is~(0,\,1), the answers must agree. Since Bob is already committed to answering~1 when asked 1, Alice must likewise answer~1 when asked 0. Continuing this line of reasoning, if the questions are~(0,\,0), they must again agree on their answers. Alice is now committed to writing~1 when asked~0, so Bob must also write~1 on question~0.

The trap is sprung in the final scenario,~(1,\,0). Alice is asked~1, and, by her prior commitment in the~(1,\,1) case, she must answer~0. Bob is asked~0, to which he is committed to answer 1, as established in the~(0,\,0) case. Following this chain of commitments, Alice and Bob are forced to return different answers, yet the rules for~(1,\,0) demand that they be the same. They lose.

This is not a failure of cleverness. By forcing a win on three of the four possible question pairs, you necessarily guarantee a loss on the fourth, leaving the win rate capped at 75\%.

Surely, there are more ways to play \emph{Coordinated Hidden Strategy Heroes}. For instance, what if the players take a more relaxed approach and answer at random? Doing so would yield only a~50\% win rate: for any question pair, whatever Alice replies, Bob has only a~50\% chance of fulfilling the rule by chance.

And what if our players mix their strategies? Perhaps they agree that in the first half of the rounds, they follow the \emph{All-Zero Strategy}, and for the remainder, they switch to the strategy that secures the~(1,\,1) scenario. Or perhaps they carry a book of random numbers to switch between plans unpredictably. 
Mixing strategies cannot help. A weighted average of strategies that each wins at most 75\% of the time still wins at most 75\% of the time.

It is unavoidable: as long as their answers are determined by parameters shared in their common past—information and instructions written on a Strategy Card—they cannot break the 75\% ceiling.

\section{An Unsettling Observation}

Now imagine the following.

Alice and Bob play not~40 but~40 \emph{million} rounds of \emph{Coordinated Hidden Strategy Heroes}. When they reconvene on Earth to compare their lists, they do not score 30 million wins, nor anything close. 
They score~34.1 million wins, giving them a consistent win rate above 85\%.

If they scored~34 out of~40 once, we might call it luck. But when they secure an~85\% win rate across millions of rounds, luck is not a viable explanation. Moreover, examining the data more closely reveals that for each of the four possible question pairs—(0,\,0),~(0,\,1),~(1,\,0) and~(1,\,1)—Alice and Bob succeed about 85\% of the time.
They are breaking the ceiling of the Strategy Card. But \emph{how}?

Are they cheating? If so, how? Are they secretly communicating faster than light? Or did they know, in their preparatory phase, which questions would be asked?

In doing their exploit, Alice and Bob were seen to carry something. Not a notebook of instructions, but a physical object—a crystal. To the casual observer, it might have looked like sentimental jewelry or a lucky charm of some kind. When answering their questions, Alice and Bob were both seen to handle their crystal. But how do these crystals permit an otherwise impossible violation of the~75\% ceiling? Is some kind of magic at work? 

Hardly. Arthur C. Clarke once remarked~\cite{clarke_profiles_1973} that sufficiently advanced technology is indistinguishable from magic, and, as David Deutsch later clarified~\cite{deutsch_fabric_1997}, this is because there is no room for magic in a comprehensible universe. Anything that appears inexplicable—be it a conjuring trick or a violation of a Bell inequality—is merely evidence that there is something we have not yet understood. 

So these crystals are not charms, but engineered physical devices. They are rare-earth-ion-doped silicates, within which lie millions of precisely identifiable charged ions trapped in a rigid lattice. The crystal host acts as an extraordinary insulator, preserving the delicate quantum states of the ions against the noise of the outside world. Effectively, these crystals are quantum hard disks.

To play~40 million rounds of CHSH, Alice and Bob each need~40 million doped ions inside their crystals. The arrangement is a strict one-to-one correspondence: for every specific ion in Alice's crystal, there is a designated partner in Bob's. Before the players separate, these ion pairs are prepared together, initialized in a joint quantum state—an entangled state. Because each round requires its own dedicated pair, the crystals must house these prepared ion pairs, ready to be measured in sequence.

When the game starts, Alice does not look up a rule. She interacts with her crystal. Targeting the specific ion for that round, she applies an optical pulse determined by her question (0 or 1)—effectively transferring the question to the ion—and observes the flash of light it emits. If the ion fluoresces, scattering a brief, bright spark of photons, she records a 0. If the ion remains dark, she records a 1. Independently, Bob does the same at his station near Jupiter. 

When they reconvene on Earth to compare their lists, they shatter the 75\% limit. Over millions of rounds, the crystals have guided them to a consistent win rate above 85\%.

Faced with such a result, we may marvel at the power of the hardware. It opens a new technological frontier. The sheer existence of these crystals implies that the universe supports unexpectedly powerful modes of information processing—a hidden computational resource waiting to be harnessed.

Yet the utility of the tool pales in comparison to the explanatory gap it exposes. We have proven that no Strategy Card—no pre-written set of local instructions—can achieve this score. The question is unavoidable: if we cannot write a strategy for Alice and Bob to win, how does nature orchestrate the win for the ions? 

This exploit is not science fiction. In modern physics laboratories, this game is played and beaten routinely. To enforce the strict isolation the rules demand, experimental physicists do not need to send equipment to Mercury and Jupiter. They simply separate the ions or other systems across a university campus and perform the measurements with ultra-fast response times. By finishing the measurements in fractions of a microsecond, they guarantee each round is over before any light-speed signal could possibly cross the distance between the measurement stations.
The mathematical limit that caps the Strategy Card at 75\% is known as a \emph{Bell inequality}. The subsequent experimental demonstration that this ceiling can be systematically shattered stands as one of the central empirical results of modern physics. This very achievement earned John Clauser, Alain Aspect and Anton Zeilinger the 2022 Nobel Prize.

\section{Holding on to Strategy Cards}

To preserve the Strategy Card as the underlying mechanism governing quantum systems, physicists have been willing to revise some of our deepest assumptions about reality. Three broad responses have emerged—each requiring the dismantling of the structure of spacetime and rewriting the rules of causality.

\subsection{Covert Hyper-Space Hotline}
One possibility is that the ions within the crystals secretly communicate faster than light—perhaps even instantaneously. When Alice's ion is measured, it somehow informs Bob's ion which answer to produce, bridging the vast distance in little to no time at all. This postulates a hidden communication mechanism that bypasses the fundamental structure of spacetime.

Such a view demands nothing less than abandoning the core principles of relativity. While mathematical models like Bohmian mechanics exist~\cite{bohm_1952_I}, they require a preferred reference frame—effectively undoing Einstein's revolution—and no physical mechanism for such a signal has ever been identified. 

The widespread belief in instantaneous influence is largely a relic of the collapse postulate, which textbook quantum theory does not turn into a coherent physical process. It specifies no mechanism, no precise criterion for when collapse occurs, and no clear account of which physical event triggers it.
 
For proponents of nonlocality, CHSH might as well stand for a \emph{Covert Hyper-Space Hotline}.

The Nobel laureate Alain Aspect explicitly champions this view. Asserting that ``we must admit some spooky action at a distance,'' he concludes: ``I prefer to name it quantum nonlocality''~\cite{aspect_worldofquantum_2025}. 
However, the physics community adopted ``quantum nonlocality'' as the standard term not just for this ``spooky'' hypothesis, but for the empirical violation of a Bell inequality itself.
This damaging conflation achieved total linguistic capture: thousands of peer-reviewed papers now casually refer to ``quantum nonlocality'' as settled physics.
Once ``nonlocality'' is tied to an unquestionable empirical result, we are rhetorically forced into a verdict about nature's covert hotline—a verdict we will refute in three different ways in the pages to follow.

\subsection{Cosmic Hoax of Scripted History}
Another proposal is that the game was rigged from the start, as if our CHSH players already knew, during the preparatory phase, which questions they would later be asked. Since the ions are effectively the players, their Strategy Card would in fact be a Grand Strategy Card, ensuring secret coordination not only between them, but also with whatever process determines the questions.

This is the \emph{superdeterminist} move~\cite{hossenfelder_rethinking_2020}. It does not merely eliminate free choice; it outlaws counterfactual reasoning itself. Questions such as ``\emph{What would the ions have done had the settings been different?}'' are treated as meaningless, thereby disabling one of the central modes of scientific inquiry.

In the language of causality, this total lack of freedom is implemented by a common cause affecting not only the ions but also the processes that determine the questions. Such a mechanism can be made to reproduce any correlation, but only at the cost of severe fine-tuning: the universe must conspire in intricate detail to preserve the appearance of independence across every experimental design meant to guarantee it.

The burden of that fine-tuning becomes especially vivid when the settings are tied to arbitrary public \emph{past} data. Suppose that on an upcoming Bell experiment, Alice's questions are determined by the parity of the last digit of the S\&P~500 closing value on selected dates in 2023, while Bob's are defined by the parity of total social-media post counts on selected dates in 2025. Then the Grand Strategy Card guiding the ions must factor in those external human-generated facts. ``Because the closing value of the S\&P~500 on February 16\textsuperscript{th}~2023 was~4\, 090, the question on Alice's side will be~0; and because there were~15\,028\,116\,089 social media posts on November 3\textsuperscript{rd}~2025, the question on Bob's side will be~1. Therefore, you two, ions, should reply~0 and~0.''

Where is that information supposed to be stored? In which local degrees of freedom and by which mechanism? 

Superdeterminism is the modern name for this view, though the idea is hardly new. It echoes the medieval doctrine of Occasionalism, according to which fire does not burn cotton; rather, God creates the burning at the moment of contact. In both pictures, local cause and effect are only appearances sustained by a deeper script. For those who find such a conspiracy too convenient to swallow, CHSH stands for a \emph{Cosmic Hoax of Scripted History}.

\subsection{Causality Held Subject to Hindsight}

A third option that clings to the Strategy Card is to allow influences to run backward in time~\cite{price_timesymmetry_2012}. On this view, the questions asked later on Mercury and Jupiter ripple into the past—back on Earth—and help determine the Strategy Card prepared there. In their preparatory phase, the ions are effectively guided by a message coming from the future, already encoding which questions will be asked. The Strategy Card is therefore not written before the game begins; it is scribbled retroactively once the questions are asked.

This proposal suffers from the same polite invisibility as the superluminal hotline. Just as the hotline allows influence across space while forbidding any usable faster-than-light signal, retrocausality allows influence across time while forbidding any usable signal to the past. The feature is tailored exclusively to save the Strategy Card, while carefully avoiding any observable consequences that would allow us to test it.

Furthermore, unlike the closed timelike curves discovered by Gödel~\cite{godel_1949_rmp}, which arise as explicit solutions to general relativity under conditions like those of a rotating universe, this retrocausality is an ad hoc modification of quantum theory, asserted to happen in every single quantum event, everywhere, all the time. No physical mechanism is proposed for how an effect swims upstream against time and rewrites the coordination of quantum systems.

In this picture, \emph{Causality is Held Subject to Hindsight}.

\section{The False Trilemma}
Each of these proposals—the Hotline, the Hoax and the Hindsight—responds to the same pressure: Bell's theorem rules out explanations based on the Strategy Card subject to the causal structure of spacetime. To save the Card, physicists have been willing to sacrifice the structure, proposing that nature is superluminal, conspiratorial, or retrocausal.

A pervasive response is explanatory resignation. An instrumentalist reflex treats Bell tests as something to be established and then filed away. The important lesson is said to be that the inequality \emph{is} violated, not \emph{how} it is violated—except insofar as the ``how'' concerns experimental loopholes.
If Boltzmann had thought that the important lesson is that heat \emph{is} conducted, not \emph{how} it is conducted, he would not have bothered developing the kinetic theory of heat based on molecular motion.
The instrumentalist posture stifles progress by disallowing the very question that drew us to physics as children: ``\emph{How does it work?}'' 

%Yet physics does not stop at predicting what \emph{is}; it advances by asking ``\emph{how}''.
%Newton's formula for gravity was not an endpoint but a provocation that culminated in general relativity.

Before resigning ourselves to the Hotline-Hoax-Hindsight trilemma or to the silence of instrumentalism, perhaps we should question the assumption that cornered us in the first place: the Strategy Card.

By insisting on Strategy Cards, we are positing that pre-established classical information, ``hidden variables'', must determine the outcome of every measurement. It is an attempt to impose the mechanics of single-valued, non-interfering entities onto a theory that explicitly claims to be alien to them. It tries to force the round peg of quantum reality into the square hole of classical information.

But here is the crucial point: \emph{genuinely classical information does not exist at the fundamental level}. With the unrivalled success of quantum theory, the tables must be turned. It is the classical realm that demands an underlying quantum explanation, not vice versa. This is the domain of decoherence theory, which describes how the interaction with the environment stabilizes quantum systems into the robust, effectively classical structures we observe.

Bell's theorem proves that if we force classical cards into quantum systems, we are compelled to invent conspicuous mechanisms to patch the cracks. But when we take quantum mechanics seriously, applying it universally to ions, apparatuses, and observers alike, there are no cracks to patch. There is no room for hidden variables and no need to break the causal structure of spacetime.
The challenge, then, is not to supplement quantum theory, but to trust it. 

\section{The Quantum Mechanism of Local Reality}

We do not begin with the assumption of a Strategy Card. 
We begin with our best physical theory. 
Therefore, we do not rewrite the laws of physics to explain how the ions coordinate; we only apply them. 
By adopting the Heisenberg-picture formalism developed by Deutsch and Hayden~\cite{deutsch_hayden_2000}, and applying it universally—to ions, detectors, environments and observers—a local account emerges.

The CHSH game is won with no Hotline, no Hoax and no Hindsight.

\subsection{Creating Histories Separately Held}

For the sake of definiteness, consider one round of \emph{Coordinated Hidden Strategy Heroes}, played by Alice and Bob with the help of entangled ions. On Mercury, Alice directs a laser pulse dictated by the question she received. As the ion interacts with the pulse, it evolves into a superposition of possible outcomes. The superposition spreads outward through the photon field, the apparatus, and soon enough, Alice herself, who is swept up into it.

According to the unitary equations of quantum mechanics expressed in the Heisenberg picture, this interaction causes Alice to smoothly and locally evolve into a superposition of two distinct versions of herself: an Alice-who-saw-0 and an Alice-who-saw-1. Her two histories carry equal measure, 50-50, which provides the physical basis for equal probability.

This is fully local, as it happens entirely within Alice's station. While photons and other particles may radiate outward, no influence can instantaneously reach remote systems.

Meanwhile, on his station orbiting Jupiter, Bob performs his own measurement. An analogous, yet independent and strictly local process occurs. Bob foliates into Bob-who-saw-0 and Bob-who-saw-1, both with equal measure.

Here is where our intuition about a global ``now'' usually tricks us. We want to link the histories and ask: ``\emph{Which version of Bob corresponds to the Alice-who-saw-0}?''

The answer from the Heisenberg picture is radical: \emph{None yet}.

Because these events are dynamically isolated, and because Alice and Bob's descriptors in the Heisenberg picture are entirely separable, Alice's and Bob's local branchings are uncoupled. 
Alice's history has branched into two equal-measure continuations, as has Bob's, and there is no ``spooky action'' or global link aligning them. 

By measuring their ions in isolation, Alice and Bob are \emph{Creating Histories Separately Held}, waiting for a future interaction to bridge them.

\subsection{Correlations Happen Safely Home}

Crucially, to confirm the violation of Bell inequalities, Alice and Bob must further interact to produce a record of the joint outcomes. In the single-world orthodoxy, bringing systems together that are each in a definite state is merely a matter of convenience for comparison. But in unitary quantum theory, that comparison is a physical interaction—a handshake—that must be explicitly analyzed within the theory. 
As we shall see, the comparison skews the measures of the joint outcomes.

For concreteness, let us take the perspective of Alice acting as the joint record. When she travels back to Earth to compare notes, she arrives as two parallel, equal-measure trajectories: the Alice-who-saw-0 and the Alice-who-saw-1.

Consider first the Alice-who-saw-0, who holds exactly half of the total measure. Upon meeting Bob and reading his result, she branches a second time, splitting into Alice-who-saw-0-and-saw-Bob-0 and Alice-who-saw-0-and-saw-Bob-1. This second branching is not even. 
The algebraic structure preserved inside Alice's and Bob's descriptors comes into play, skewing the branching. 
If at least one CHSH question is~0, the Alice-who-saw-0-and-saw-Bob-0 takes an~85\% share of Alice-who-saw-0’s measure, while the ``losing'' Alice takes the remaining~15\%. Had the question pair been~(1,\,1), Alice-who-saw-0-and-saw-Bob-1 would have taken the~85\% share.

A symmetrical process happens to the parallel Alice-who-saw-1, whose trajectory also branches 85/15 upon reading Bob's result, with the dominant measure attributed to the CHSH-winning match.

The final result is precise: the winning output pairs carry a total measure of~85\%. The correlation does not come from prearranged coordination in the common past; it arises at the later, wholly local comparison event, when the players meet and re-branch, skewing the measure. 

When playing CHSH, \emph{Correlation Happens Safely Home}.

\subsection{Classicality Hosted by Scattered Handoffs}

At first glance, the story of local branching and later re-branching seems to clash with the standard lore about quantum theory. Quantum effects are notoriously delicate: they require isolation as they are easily washed out in warm, noisy environments. 
While that strict isolation was maintained for the ions on their initial journeys to Mercury and Jupiter, macroscopic observers afford no such luxury.
%So how could branched Alices and Bobs persist long enough to later undergo the skewed branching that yields Bell-violating statistics?
So how could the algebraic structure preserved inside Alice's and Bob's descriptors persist long enough to later drive the skewed branching that yields Bell-violating statistics?

The key is that in universal quantum mechanics, ``classical'' does not mean ``non-quantum''—it cannot. It means robust. Classicality is the regime in which information is copied so widely that it gives rise to stable, autonomous histories. 

When Alice measures her ion, the outcome is copied into her detector, into her memory, into scattered photons, and into countless other nearby degrees of freedom. 
This massive redundancy weaves a web of witnesses that stabilizes the record; once copied so widely, the~``0'' and~``1'' alternatives can no longer be recombined via quantum interference. 
%Each is supported by a macroscopic ledger that continually reinforces it. 
Therefore, classicality is not the absence of quantum structure; it is a particular quantum structure—one built from redundant witnesses.

The same logic explains what we call classical communication. In a fully quantum universe, there is no special, rigid ``classical channel''. Physically, what travels inside a telephone wire is not an abstract bit, but a record passed along by a chain of local quantum interactions. 
%Instead, information is transmitted by a chain of local handoffs. 
Each system couples to the next in a scattered relay of redundant copies.
In CHSH, like elsewhere, 
%classicality is not a layer imposed on top of quantum mechanics. It is the emergent regime in which records persist because they are copied and handed off again and again. 
\emph{Classicality is Hosted by Scattered Handoffs}.

And now the crucial point for Bell and the CHSH game: the very processes that make records classical—redundant copying and long chains of handoffs—do not obstruct the 85/15 branching at comparison. 
%Those processes do not need to preserve a delicate, isolated quantum superposition across interplanetary space; that burden belonged to the isolated ions on their initial journeys to Mercury and Jupiter. What must survive between Alice's measurement and her comparison is far more mundane: a stable record of her measurement outcome. 
When Alice meets Bob and reads his result, classicality does not flatten the \emph{quantum} structure that gives the handshake its skew.

\section{Conclusion}

Bell's theorem was supposed to be the end of the local story. 
We inherited a verdict that no account in spacetime could explain quantum correlations—that they must arise ``from outside space–time''.
But Bell's inequality does not rule out local reality; it rules out a fundamentally classical, single-history, local universe.
Yet those who considered quantum mechanics to be universal already knew that.

When every physical system in the CHSH game is analyzed entirely within the unitary Heisenberg picture, nothing leaps across the void.
Histories are created locally, held separately, and only later brought into contact, where they jointly evolve to forge the correlations.
% The correlations do not arise from distant coordination; they arise from what happens when autonomous records meet and jointly evolve. 
%By trusting our best physical theory, we dispense with quantum magic. 
%
Local reality survives the Bell test intact, as \emph{Colliding Histories Skew Handshakes.}
%Again, locality is restored.

\subsection*{Acknowledgements}

I am grateful to Maxime Desalle, Hervé Eulacia, André Garcia and Samuel Kuypers for helpful comments, and I am especially thankful to Logan Chipkin for his meticulous review of an earlier draft.
I also wish to thank the Conjecture Institute for welcoming me as a fellow and for its intellectual support.
%I am grateful to Maxime Desalle, Hervé Eulacia, Samuel Kuypers, and André Garcia. I owe special thanks to Logan Chipkin for his exceptionally careful feedback on earlier versions of this manuscript. I also wish to thank the Conjecture Laboratory for welcoming me as a fellow and for its intellectual support.
%
%I am also grateful to XXX, and I thank YYY as well for providing...
%%
%I wish to thank ZZZ for their invaluable support.
%%
%Finally, I also wish to thank the Conjecture Institute for welcoming me as a fellow and for its intellectual support, particularly through the careful feedback of Logan Chipkin.

This work was supported by the Mitacs Elevate postdoctoral fellowship in partnership with Bbox Digital.

\end{document}